\documentclass[prb,preprint]{revtex4}

\pdfoutput=1

\usepackage{graphicx}
\usepackage{latexsym}
\usepackage{amssymb}
\usepackage{amsmath} 

\begin{document}


\title{Closing of the pseudogap in $\mathbf{La_{2-z-x}Nd_zSr_xCuO_4}$  ($\mathbf{z = 0, \;0.4}$)}


\medskip 

\date{October 10, 2021} \bigskip

\author{Manfred Bucher \\}
\affiliation{\text{\textnormal{Physics Department, California State University,}} \textnormal{Fresno,}
\textnormal{Fresno, California 93740-8031} \\}

\begin{abstract}
Doping of $La_2CuO_4$ and $La_{1.6}Nd_{0.4}CuO_4$ with $Sr$ gives rise to holes that locate pairwise at lattice-site $O$ atoms. Such $O$ atoms reside as lattice defects in the $CuO_2$ planes if the doping level $x$ is below a watershed value, $x < \hat{x}$, but also in the bracketing $LaO$ layers if $x > \hat{x}$. The $O$ atoms form a 2D charge order of incommensurability $q_{CO}^{CuO_2}(x)$ and $q_{CO}^{LaO}(x)$. At the doping $x^*$ (quantum critical point) that causes the closing of the pseudogap at $T=0$, $q_{CO}^{CuO_2}(x^*) = 2q_{CO}^{LaO}(x^*)$ holds.

\end{abstract}

\maketitle

\section{CHARGE ORDER IN $\mathbf{La_{2-x}Sr_xCuO_4}$
}

When $La_2CuO_4$ is doped with strontium, substitution of $La \rightarrow La^{3+} + 3 e^-$ by $Sr \rightarrow Sr^{2+} + 2 e^-$ causes a lack of electrons, or equivalently, a creation of holes. Coulomb repulsion spreads the holes, which then attach pairwise to oxygen in the $CuO_2$ planes, $O^{2-} + \;2e^+ \rightarrow O$. There the $O$ atoms form a charge order (CO) by a 2D superlattice. The reciprocal of the superlattice spacing, $A_0(x) \simeq B_0(x)$, is called the incommensurability,\cite{1} 
\begin{equation}
\frac{1}{A_0(x)}\equiv q_{CO}^{CuO_2}(x)= \frac{\Omega^{\pm}}{{2}}\sqrt {x - \tilde{p}}\;,\;\;\;\;\;x < \hat{x}  \; .
\end{equation}
In the medium doping range, $0.10 \preccurlyeq x \preccurlyeq  0.15$, the charge order manifests as unidirectional stripes, at higher doping as nematicity.
The formula, in reciprocal lattice units (r.l.u.), is valid for doping up to a ``watershed'' concentration $\hat{x}$, which depends on the species of doping. The  stripe-orientation factor is $\Omega^{+}=\sqrt{2}$ for $x > x_6 = 2/6^2  \simeq 0.056$ when stripes are parallel to the $a$ or $b$ axis, but $\Omega^{-} = 1$ for $x < x_6$ when stripes are diagonal. The offset value $\tilde{p}$ under the radical is the hole concentration necessary to keep three-dimensional antiferromagnetism (3D-AFM) suppressed. Their skirmisher task keeps those ``suppressor holes'' from participating in charge order. 

Also because of Coulomb repulsion, the doped-hole concentration saturates in the $CuO_2$ planes at $\hat{x}$, causing the square-root curve from Eq.(1) to level off, with increased doping, to a constant plateau (depending on dopant and co-dopant species),\cite{1}
\begin{equation}
    q_{CO}^{CuO_2}(x) = \frac{\sqrt{2}}{2} \sqrt{\hat{x} - \tilde{p}} \;,\;\;\;\; x \ge \hat{x} \;,
\end{equation}
(see Fig. 1). Additional doped holes then overflow to the $LaO$ layers that sandwich the $CuO_2$ planes, where they also reside pairwise in (apical) $O$ atoms.
Again, Coulomb repulsion spreads the double holes to a planar superlattice of lattice-defect $O$ atoms in \textit{each} $LaO$ layer, with attending charge order of incommensurability\cite{1}
\begin{equation}
    q_{CO}^{LaO}(x)  = \frac{\sqrt{2}}{2}\sqrt {\frac{x - \hat{x}}{2} }\;,\;\;\;\;\;  x \ge \hat{x} \; .
\end{equation}
The denominator 2 under the radical is due to the two $LaO$ layers per unit cell.

The $Sr$ doping that causes the closing of the pseudogap at $T=0$ is denoted as $x^*$.  It is widely regarded a quantum critical point.
Inserting the observed data of $\hat{x}$, $\tilde{p}$, and $x^*$ of $La_{2-x}Sr_xCuO_4$ from Table I into Eqs. (2, 3)  gives incommensurabilities  
\begin{equation}
2q_{CO}^{LaO}(x^*)= q_{CO}^{CuO_2}(\hat{x})= q_{CO}^{CuO_2}(x^*) \;. 
\end{equation}
For the sake of graphic visualization we combine the defect-charge order of the upper $LaO$ layer ($\overline{LaO}$) and lower $LaO$ layer ($\underline{LaO}$), staggered by half a superlattice spacing, $\frac{1}{2}A_0^{LaO} =\frac{1}{2}/q_{CO}^{LaO}$ (see Fig. 2), such that \textit{together} they form a charge order of incommensurability 
\begin{equation}
q_{CO}^{\overline{LaO}+\underline{LaO}}(x) = 2q_{CO}^{LaO}(x) = \sqrt{x-\hat{x}} \;.
\end{equation}
\bigskip

\includegraphics[width=6.5in]{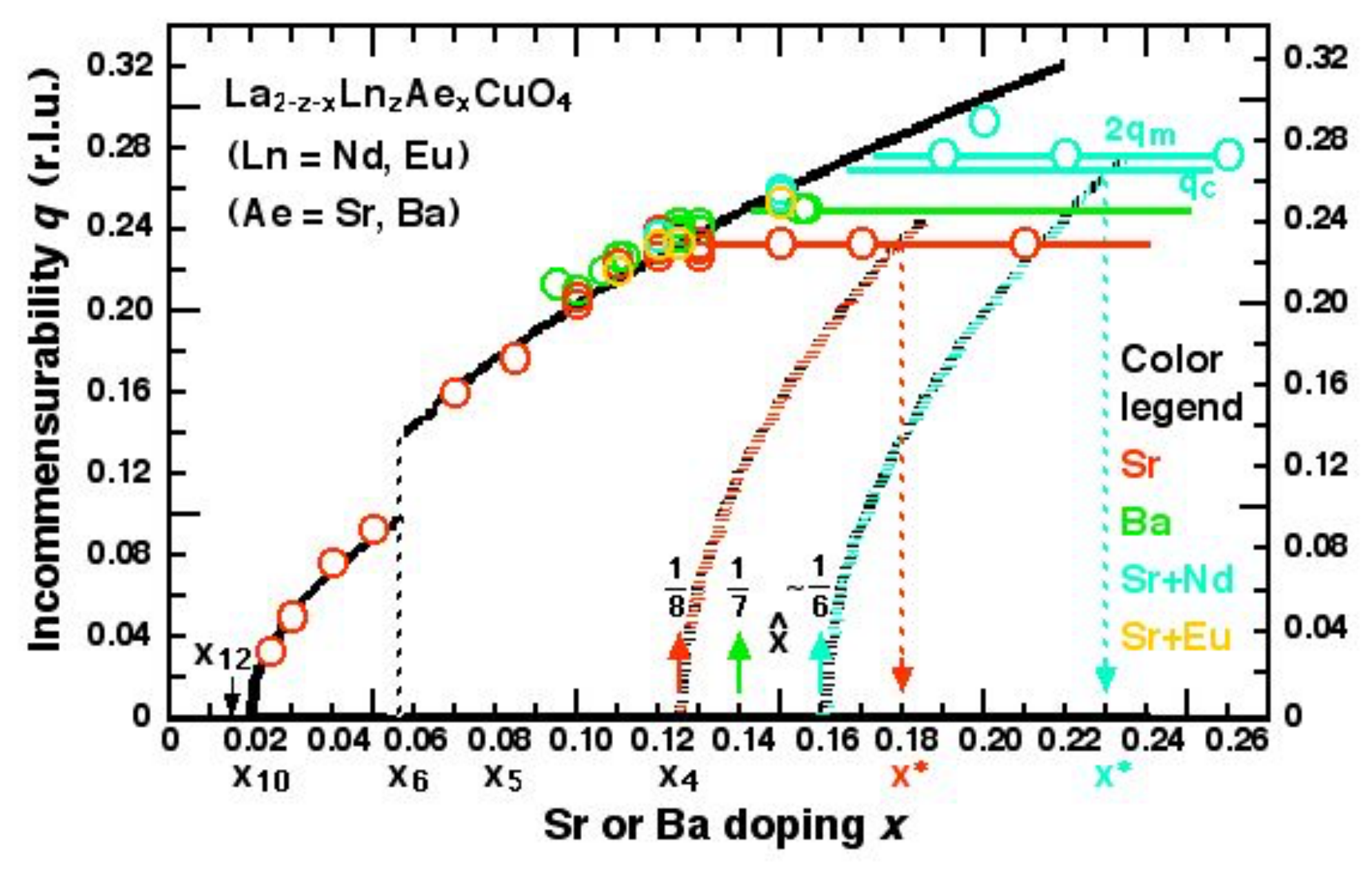}  \footnotesize 

\noindent FIG. 1. Incommensurability of $O$ superlattice charge order, $q = q_c$, and $O$ superlattice magnetization, $q = 2q_m$, in the $CuO_2$ planes (solid curves and horizontal lines) and in the $LaO$ layers (hatched curves) of $La_{2-z-x}Ln_zAe_{x}CuO_{4}$ ($Ln = Nd, Eu; z = 0, 0.4, 0.2$) due to doping with $Ae = Sr$ or $Ba$. Commensurate doping concentrations are denoted by $x_n \equiv 2/n^2$. Circles show data from X-ray diffraction and scattering or neutron scattering. For $La_{1.6}Nd_{0.4}Sr_{x}CuO_{4}$ the $2q_m$ data are experimental, the horizontal $q_c$ line is inferred (see text).
The broken black curve is a graph of Eq. (1), calculated with a constant offset value, $\tilde{p} = x_{10} = 0.02$. Its discontinuity at $x_6 \simeq 0.056$ is caused by a change of stripe orientation, relative to the planar crystal axes, from diagonal for $x<x_6$ to parallel for $x>x_6$. 
The solid (colored) horizontal lines are graphs of Eq. (2) and the hatched colored curves are graphs of $2q_c^{LaO}(x)$, Eqs. (3, 5). The condition for closing of the pseudogap, Eq. (4), is illustrated by the dashed down arrows.
\normalsize

\begin{table}[ht]
\footnotesize \caption{Watershed concentration $\hat{x}$ of doped $Sr$ where the square-root curve of charge-order incommensurability in the $CuO_2$ plane, $q_{CO}^{CuO_2}(x)$, levels off; concentration of 3D-AFM suppressor holes $\tilde{p}$; and $Sr$ doping $x^*$ at the closing of the pseudogap at $T=0.$ The charge-order incommensurability in \textit{each} $LaO$ layer is denoted as $q_{CO}^{LaO}$.} \normalsize
\begin{tabular}{|p{4cm}|p{1.8cm}|p{2cm}|p{1.7cm}|p{1.8cm}|p{2cm}|p{2cm}|}
 \hline  \hline
Compound &$\;\;\;\;\;\;\hat{x}$&$\;\;\;\;\;\;\tilde{p}$&$\;\;\;\;x^*$&$\; q_{CO}^{CuO_2}(\hat{x}) $&$\; q_{CO}^{CuO_2}(x^*) $&$\; 2q_{CO}^{LaO}(x^*) $\\
 \hline  \hline
$La_{2-x}Sr_xCuO_4$&$\;\;\;\;0.125 $&$\;\;\;0.015 $&$\;\;\;0.18 $&$\;\;\;\mathbf{0.235} $&$\;\;\;\mathbf{0.235} $&$\;\;\;\mathbf{0.235} $\\ \hline
$La_{1.6-x}Nd_{0.4}Sr_xCuO_4$&$\;\;\;\;0.16 $&$\;\;\;0.019 $&$\;\;\;0.23$&$\;\;\;\mathbf{0.266}$&$\;\;\;\mathbf{0.266}$&$\;\;\;\mathbf{0.266}$\\ \hline
 \hline
\end{tabular}
\label{table:1}
\end{table}

\includegraphics[width=6in]{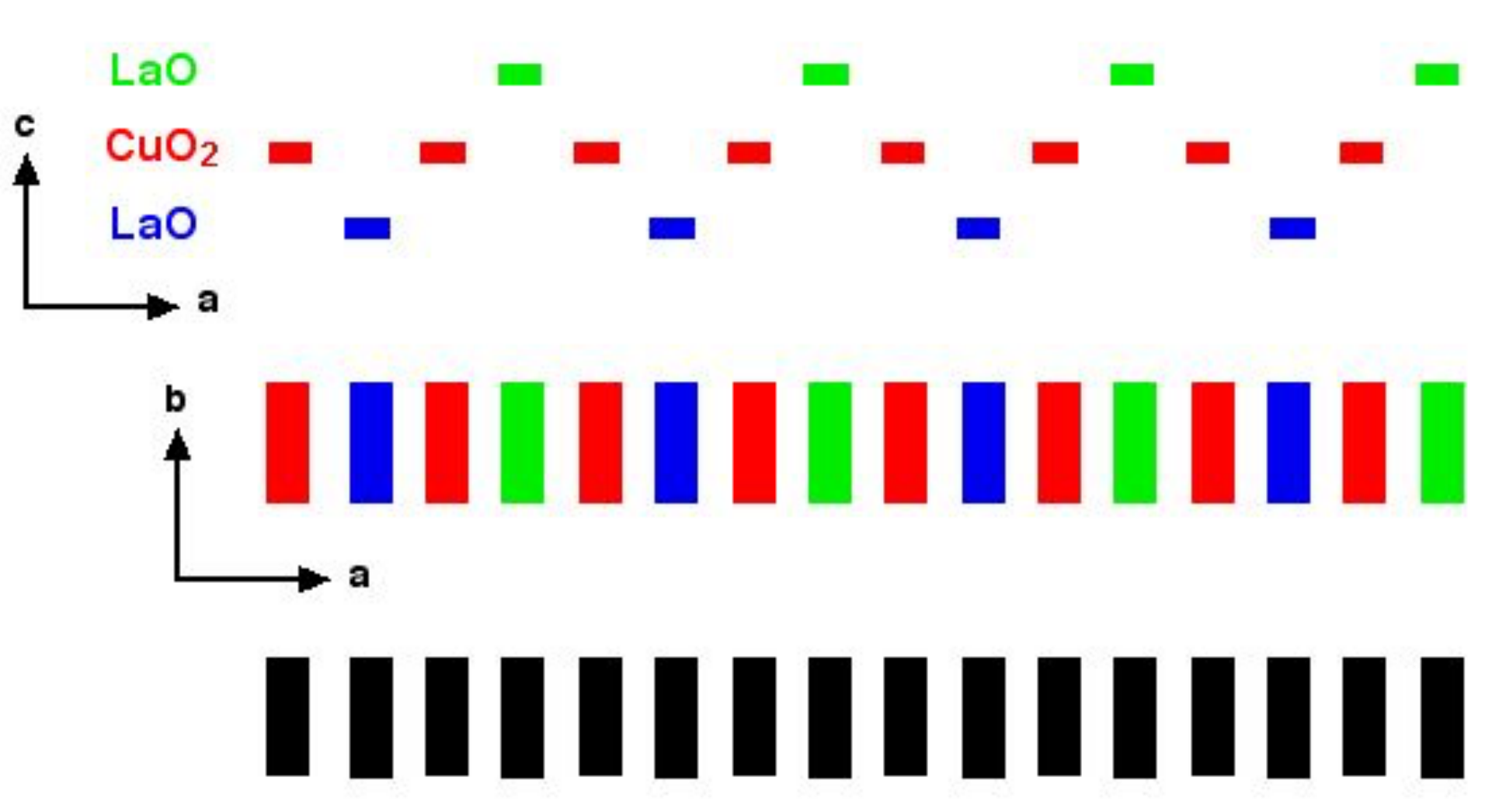}  \footnotesize 
\noindent FIG. 2. Cartoon illustrating charge-order stripes in the $CuO_2$ plane (red), upper $LaO$ layer ($\overline{LaO}$, green), and lower $LaO$ layer ($\underline{LaO}$, blue) in side view (top part) and down view (center and bottom part). At doping $x^*$, shown here, \textit{each} $LaO$ layer has an incommensurability $q_{CO}^{LaO}(x^*)= \frac{1}{2}q_{CO}^{CuO_2}(x^*)$. When their stripes are staggered and combined, both bracketing $LaO$ layers \textit{together} have the same incommensurability as the bracketed $CuO_2$ plane, $q_{CO}^{\overline{LaO}+\underline{LaO}}(x^*)= q_{CO}^{CuO_2}(x^*)$. The stripes of the combined $LaO$ layers are staggered against those of the $CuO_2$ plane such that they interlace.
\normalsize

\bigskip \bigskip  

Solving Eq. (4) for $x^*$ relates the $Sr$-doping that causes closing of the pseudogap (at $T=0$) with the watershed doping $\hat{x}$ and the offset concentration $\tilde{p}$,
\begin{equation}
   x^* = \frac{3\hat{x} - \tilde{p}}{2} \;.
\end{equation}


\section{SELECTION OF DATA}
The condition of Eq. (6) for the three parameters $\hat{x}$, $\tilde{p}$ and $x^*$ 
is satisfied, within error bars, for the data of $La_{2-x}Sr_xCuO_4$, listed in Table I. 
The offset value $\tilde{p} = 0.015 \pm 0.002$ is confirmed, through Eq. (1), by many measurements of charge-order and magnetization stripes in $La_{2-x}Sr_xCuO_4$ near $x = 0.125 = 1/8$, cited in Ref. 1. Visually, it can be inferred from the accumulation of data points \textit{slightly above} the square-root curve in Fig. 1 (drawn with constant $\tilde{p} = x_{10} = 0.02$) for $x > 0.09$ but coincident with a slightly upshifted curve in that range (not shown) when $\tilde{p} = 0.015$ is used. The value of the watershed doping in Table I, $\hat{x} = 0.125$, is obtained by Eq. (2) from a \textit{constant} charge-order incommensurability, $q_{CO}^{CuO_2} = 0.235, x \ge \hat{x}$,  being within error bars of $q_{CO}^{CuO_2} = 0.235 \pm 0.005$ r.l.u., observed with non-resonant hard X-rays, and $q_{CO}^{CuO_2} = 0.231 \pm 0.005$ r.l.u., observed with resonant inelastic soft X-ray spectroscopy.\cite {2,3} The doping level where the pseudogap closes (at $T=0$), $x^* = 0.18 \pm 0.01$, has been determined with resistivity and Nernst-effect measurements.\cite{4,5}

Turning to the data for $La_{1.6-x}Nd_{0.4}Sr_xCuO_4$, an offset value $\tilde{p} = 0.019$ is obtained with Eq. (1) from $q_{CO}^{CuO_2} = 0.23 \pm  0.005$ r.l.u. of $La_{1.475}Nd_{0.4}Sr_{0.125}CuO_4$.\cite{6} 
Measurements of resistivity, Hall effect, Nernst effect, and thermoelectric power give $x^* = 0.23 \pm 0.01$.\cite{7,8,9} A \textit{constant} incommensurability of magnetization stripes, $q_M^{CuO_2} = 0.139 \pm 0.02$ r.l.u. has been observed, at temperature $T = 1.5$ K and doping $x = 0.19, 0.24, 0.26$, with neutron scattering.\cite{10} If the relationship between charge order and magnetic order,
\begin{equation}
q_{CO}^{CuO_2}(x) = 2 q_M^{CuO_2}(x) \;,
\end{equation}
holds, as for $La_{2-x}Ae_xCuO_4$ ($Ae = Sr, Ba$), then $q_{CO}^{CuO_2}=0.278$, together with Eq. (2), would give a value of $\hat{x}=0.17 \pm 0.005$. Although close, the values of $\tilde{p}$, $x^*$ and $\hat{x}$ thusly obtained do \textit{not} satisfy the condition of Eq. (6) \textit{within error bars}. What could be the reason?

It is known that the magnetic moment of the co-dopant, $\mathbf{m}(Nd^{3+})\ne 0$ --- as opposed to $\mathbf{m}(La^{3+}) = 0$ --- couples with the magnetic moment of the host $Cu^{2+}$ ions, $\mathbf{m}(Cu^{2+})\ne 0$, at temperatures $T < 5$ K.\cite{10,11} This is the reason why magnetization stripes are observable in $La_{1.6-x}Nd_{0.4}Sr_xCuO_4$ at such high doping, in contrast to $La_{2-x}Ae_xCuO_4$. The $\mathbf{m}(Nd^{3+})$ moments may also couple with the magnetic moments of the lattice-defect $O$ atoms that give rise to the magnetization stripes, $\mathbf{m}(O)\ne 0$, such that the relation of Eq. (7) is slightly violated in $La_{1.6-x}Nd_{0.4}Sr_xCuO_4$. If we instead assume that Eq. (6) is valid, we obtain $\hat{x} = (2x^* + \tilde{p})/3 = 0.16$, listed in Table I, and a charge-order incommensurability $q_{CO}^{CuO_2}(\hat{x}) = 0.266$ instead of $2q_M^{CuO_2}=0.278$ (see Fig. 1), in slight violation of Eq. (7).

Besides $La_{2-x}Sr_xCuO_4$ and $La_{1.6-x}Nd_{0.4}Sr_xCuO_4$, treated here, there remain two more heterovalent-metal doped cuprates with high-$T_c$ superconductivity: 
$La_{2-x}Ba_xCuO_4$ and 
$La_{1.8-x}Eu_{0.2}Sr_xCuO_4$.
Because of difficulties in crystal growth, few data beyond $x = 0.125$ have been observed for $La_{2-x}Ba_xCuO_4$. With $\hat{x} = 0.14$ and $\tilde{p} = 0.015$,\cite{12} determined by Eqs. (2, 1), a prediction of $x^* = 0.20$ can be made with Eq. (6), listed in Table II. Likewise, with $\tilde{p} = 0.015$ and $x^* = 0.23$ of $La_{1.8-x}Eu_{0.2}Sr_xCuO_4$,\cite{13,8} the watershed doping for that compound is predicted as $\hat{x} = 0.16$ and the constant incommensurability of charge order, $q_{CO}^{CuO_2} = 0.266$---both in agreement with 
$La_{1.6-x}Nd_{0.4}Sr_xCuO_4$.

\begin{table}[ht]
 
\begin{tabular}{|p{4cm}|p{1.8cm}|p{2cm}|p{1.8cm}|p{1.7cm}|}
 \hline  \hline
Compound &$\;\;\;\;\;\;\hat{x}$&$\; q_{CO}^{CuO_2}(\hat{x}) $&$\;\;\;\;\;\;\tilde{p}$&$\;\;\;\;x^*$\\
 \hline  \hline
$La_{2-x}Ba_xCuO_4$&$\;\;\;\;0.14 $&$\;\;\;0.25 $&$\;\;\;0.015 $&$\;\;\;\mathbf{0.20} $\\ \hline
$La_{1.8-x}Eu_{0.2}Sr_xCuO_4$&$\;\;\;\;\mathbf{0.16} $&$\;\;\;\mathbf{0.266} $&$\;\;\;0.015 $&$\;\;\;0.23$\\ \hline
 \hline
\end{tabular}
\caption{Same notations as in Table I. Predictions with Eqs. (6, 2) are marked bold.}
\label{table:2} \end{table}

\section{SHIFT OF $\mathbf{x^*}$ BY $\mathbf{Nd}$ CO-DOPING}

A question often asked is: Why is the high-doping end of the pseudogap phase different for $La_{2-x}Sr_xCuO_4$ and $La_{1.6-x}Nd_{0.4}Sr_xCuO_4$, with $x^* = 0.18$ for the former but $x^* = 0.23$ for the latter?\cite{14} If $x^*$ is affected by the watershed doping $\hat{x}$, as can be inferred from Eq. (4), then the question is better rephrased as: Why is $\hat{x} = 0.125$ in $La_{2-x}Sr_xCuO_4$  but $\hat{x} = 0.16$ in $La_{1.6-x}Nd_{0.4}Sr_xCuO_4$?

It is likely that the difference is due both to different magnetic moments of co-dopant and host ions, $\mathbf{m}(Nd^{3+})\ne \mathbf{m}(La^{3+}) = 0$, and to different ions sizes, $r(Nd^{3+}) = 1.26$ \AA $
\linebreak
<$ $r(La^{3+})= 1.30$ \AA. The former gives rise to repulsive magnetic interaction between $\mathbf{m}(Nd^{3+})$ and $\mathbf{m}(O)$ moments in the $(La/Nd)O$ layers. The latter causes a slightly smaller lattice  constant $c_0$ of the $Nd$-codoped compound, and consequently a slightly smaller unit-cell volume $v_0$ (see Table III). Conceptually, this makes it harder for doped holes to enter the $(La/Nd)O$ layers and form lattice-defect atoms,
$O^{2-} +\; 2e^+ \rightarrow O$. The necessary higher internal pressure is achieved by Coulomb repulsion of doped holes in the $CuO_2$ planes at a higher density, to wit by $\hat{x} = 0.125 \rightarrow 0.16$.

\begin{table}[ht]
\begin{tabular}{|p{3.9cm}||p{1.3cm}|p{1.3cm}|p{1.3cm}|p{1.4cm}||p{1.5cm}|p{1.5cm}|p{1.5cm}|p{1.5cm}|}
 \hline  \hline
$\;\;\;$Doping $\rightarrow$ &$\;x = 0$&$\;x = 0$&$\;x = 0$&$\;x = 0$&$x = 0.12$&$x = 0.12$&$x = 0.12$&$x = 0.12$\\
\hline
$\downarrow$ Compound &$\;a_0$ (\AA)&$\;b_0$ (\AA)&$\;c_0$ (\AA)&$\;v_0$ (\AA$^3$)&$\;\;a_0$ (\AA)&$\;\;b_0$ (\AA)&$\;\;c_0$ (\AA)&$\;v_0$ (\AA$^3$)\\
 \hline  \hline

$La_{2-x}Sr_xCuO_4$&$\;\;\;5.34 $&$\;\;\;5.43 $&$\;\;13.12 $&$\;\;\;\;380 $&$\;\;\;5.33 $&$\;\;\;5.36 $&$\;\;13.18 $&$\;\;\;\;377 $\\ \hline
$La_{1.6-x}Nd_{0.4}Sr_xCuO_4$&$\;\;\;5.34 $&$\;\;\;5.40 $&$\;\;13.03 $&$\;\;\;\;376 $&$\;\;\;5.32 $&$\;\;\;5.32 $&$\;\;13.13 $&$\;\;\;\;372 $\\ \hline
 \hline
\end{tabular}
\caption{Lattice constants $a_0, b_0, c_0$  and unit-cell volume, $v_0=a_0b_0c_0$, of $La_{2-x}Sr_xCuO_4$ and $La_{1.6-x}Nd_{0.4}Sr_xCuO_4$ without $Sr$ doping, $x=0$, and at $x=0.12$ (Refs. 15, 16).} 
\label{table:3}
\end{table}

\section{$\mathbf{Sr}$-DOPING BEYOND $\mathbf{x^*}$}
What happens with further $Sr$-doping of the lanthanum/neodymium cuprates, $x > x^*$? The constancy of $q_{CO}^{CuO_2}(x)$ for $x>x^*$, shown in Fig. 1, implies that none of the additional doped holes generate more defect $O$ atoms in the $CuO_2$ plane. Neither do they generate more defect $O$ atoms in the bracketing $(La/Sr)O$ layers --- otherwise the condition of Eq. (4) would fail and the pseudogap would open again. This leaves as the most probable scenario that the additional doped holes will be \textit{delocalized}. The assessment agrees with the notion of a Fermi-liquid phase beyond the quantum critical point, $x^*$, where the half-occupied $ Cu \; d_{x^2 - y^2}$ orbitals---being localized in the Mott insulator ($x < 0.02$) and in the pseudogap phase ($0.02 < x < x^*$), except when part of Fermi arcs---are now delocalized.

A transition of hole density (per Cu atom) from 
\begin{equation}
    n^+ \simeq x \;\;\;\;\;\;\;\;\rightarrow\;\;\;\;\;\;\;n^+ \simeq 1 + x
\end{equation}
\begin{align*} 
at\;\;\;\;\;\;\;\;\;\;\;\;\;\;\;\;\;\;\;x \;\;\;\;\;<\;\;\;x^*\;\;\;<\;\;\;\;x\;\;\;\;\;\;\;\;\;\;\;\;\;\;\;\;\;\;\;\;\;\;\;\;\;\;
\end{align*}
has been shown by measurements of electrical resistivity $\rho$, Hall coefficient $R_H$, and Seebeck coefficient $S$ (thermopower) in $La_{1.6-x}Nd_{0.4}Sr_xCuO_4$.\cite{7,9} It can be interpreted as a massive charge delocalization.
From transport measurements in $La_{1.6-x}Nd_{0.4}Sr_xCuO_4$ under pressure, a doping condition for a Lifshitz transition, $x^* \le x^{\between}$, was obtained.\cite{17}
(A Lifshitz transition is a change of the Fermi surface between hole-like and electron-like, centered in cuprates at the M = ($\frac{1}{2},\frac{1}{2}$) and $\Gamma = 0$ point of the Brillouin zone, respectively.)
At a Lifshitz 

\includegraphics[width=5.82in]{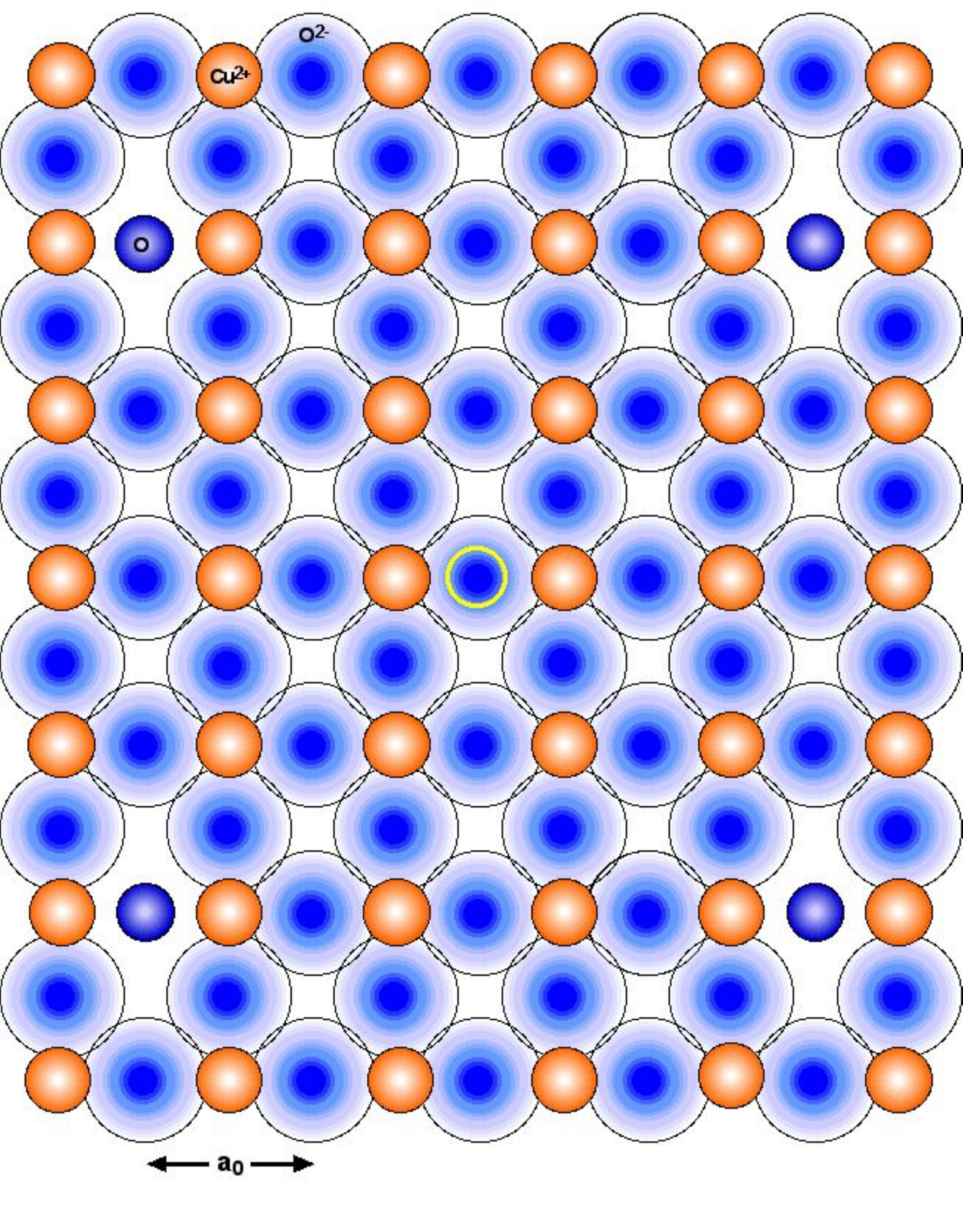}  \footnotesize 

\noindent FIG. 3. Positions of host-lattice ions and lattice-defect $O$ atoms in the $CuO_2$ plane, showing a unit cell of a \textit{commensurate} $O$ superlattice of period-$4$, $A_0 = 4 a_0$. It is roughly comparable to the incommensurate superlattice with $A_0 = 4.25 a_0$ when the pseudogap closes in $La_{2-x}Sr_xCuO_4$ or $A_0 = 3.75 a_0$ for  $La_{1.6-x}Nd_{0.4}Sr_xCuO_4$. A period-8 superlattice is formed by $O$ atoms in each bracketing $(La/Sr)O$ layer --- one such $O$ atom is indicated by the yellow circle. The host lattice is shown in its \textit{unrelaxed} position. An extended and exclusive display of all three superlattices is shown in Fig. 4. 
\normalsize

\noindent transition the carrier density can be expected to change from holes to electrons,\cite{18} 
\begin{equation}
    n^+ \simeq x \;\;\;\;\;\;\;\;\rightarrow\;\;\;\;\;\;\;n^- \simeq 1 - x
\end{equation}
\begin{align*} 
at\;\;\;\;\;\;\;\;\;\;\;\;\;\;\;\;\;\;\;x \;\;\;\;\;<\;\;\;x^\between\;\;\;<\;\;\;\;x\;.\;\;\;\;\;\;\;\;\;\;\;\;\;\;\;\;\;\;\;\;\;\;\;\;\;
\end{align*}
It has been pointed out that the charge-density relation (8), if obtained from Hall coefficient $R_H$ and magnetoresistivity $\rho_{xx}$, holds only in the high-magnetic-field limit.\cite{18} This may explain possible discrepancies between relations (8) and (9). Another issue, not completely resolved, is the Seebeck coefficient, $S > 0$, being at odds with the electron-like Fermi surface in the highly overdoped range.\cite{19,20}

In conclusion, no change of charge order is observed beyond closing of the pseudogap. Instead, a significant change of the band structure can be expected, along with a massive reconstruction of the Fermi surface---details of which await clarification---and entrance to the Fermi-liquid phase.

\appendix

\section{APPROXIMATE TREATMENT OF PSEUDOGAP CLOSING}
In view of the symmetric interlacing of $O$-atom superlattices in the $CuO_2$ plane and the bracketing $(La/Sr)O$ layers, a simplified, albeit approximate, derivation of pseudogap closing can be obtained with two approximations: (i) The circumstance that the doped-hole saturated $CuO_2$ plane is close to a period-4 superlattice of defect $O$ atoms, $\hat{x} \approx x_4^{CuO_2}$ (see Fig. 3).
(ii) Use of \textit{commensurate} doping concentrations for period-$n$ superlattices, $x_n = 2/n^2$, consistent with Eq. (1) for neglected offset, $\tilde{p} = 0$.
Holes doped beyond the watershed value $\hat{x}$ settle in the $(La/Sr)O$ layers. To fill \textit{one} such layer with a period-$8$ superlattice requires additional doping by $x_8^{LaO_2}  = 2/8^2 = 1/32 $ (yellow or green circles in Fig. 4). To fill both requires $2x_8^{LaO_2} = 2/32 = 0.0625$. As Fig. 4 shows, the $O$ atoms of the $CuO_2$ plane interlace with the projection of the $O$ atoms of the $(La/Sr)O$ layers. The total hole doping to achieve the symmetric configuration approximately is $x^{tot} = \hat{x} + 2x_8^{LaO_2} \approx x^*$ (see Table IV).

\begin{table}[ht]
\footnotesize \caption{Watershed concentration $\hat{x}$ of doped $Sr$, hole concentration in both $(La/Sr)O$ layers, $2x_8^{LaO_2}$, total hole concentration of the approximation, $x^{tot}$, and quantum critical point $x^*$.} \normalsize
\begin{tabular}{|p{4cm}|p{1.8cm}|p{2cm}|p{1.7cm}|p{1.8cm}|p{2cm}|p{2cm}|p{2cm}|}
 \hline  \hline
Compound &$\;\;\;\;\;\;\hat{x}$&$\;\;\;2x_8^{LaO_2}$&$\;\;\;\;x^{tot}$&$\;\;\;\;x^*$\\
 \hline  \hline
$La_{2-x}Sr_xCuO_4$&$\;\;\;\;0.125 $&$\;\;\;0.063 $&$\;\;\;0.188 $&$\;\;\;0.18 $\\ \hline
$La_{1.6-x}Nd_{0.4}Sr_xCuO_4$&$\;\;\;\;0.16 $&$\;\;\;0.063 $&$\;\;\;0.223 $&$\;\;\;0.23$\\ \hline
 \hline
\end{tabular}
\label{table:4}
\end{table}

\includegraphics[width=5.7in]{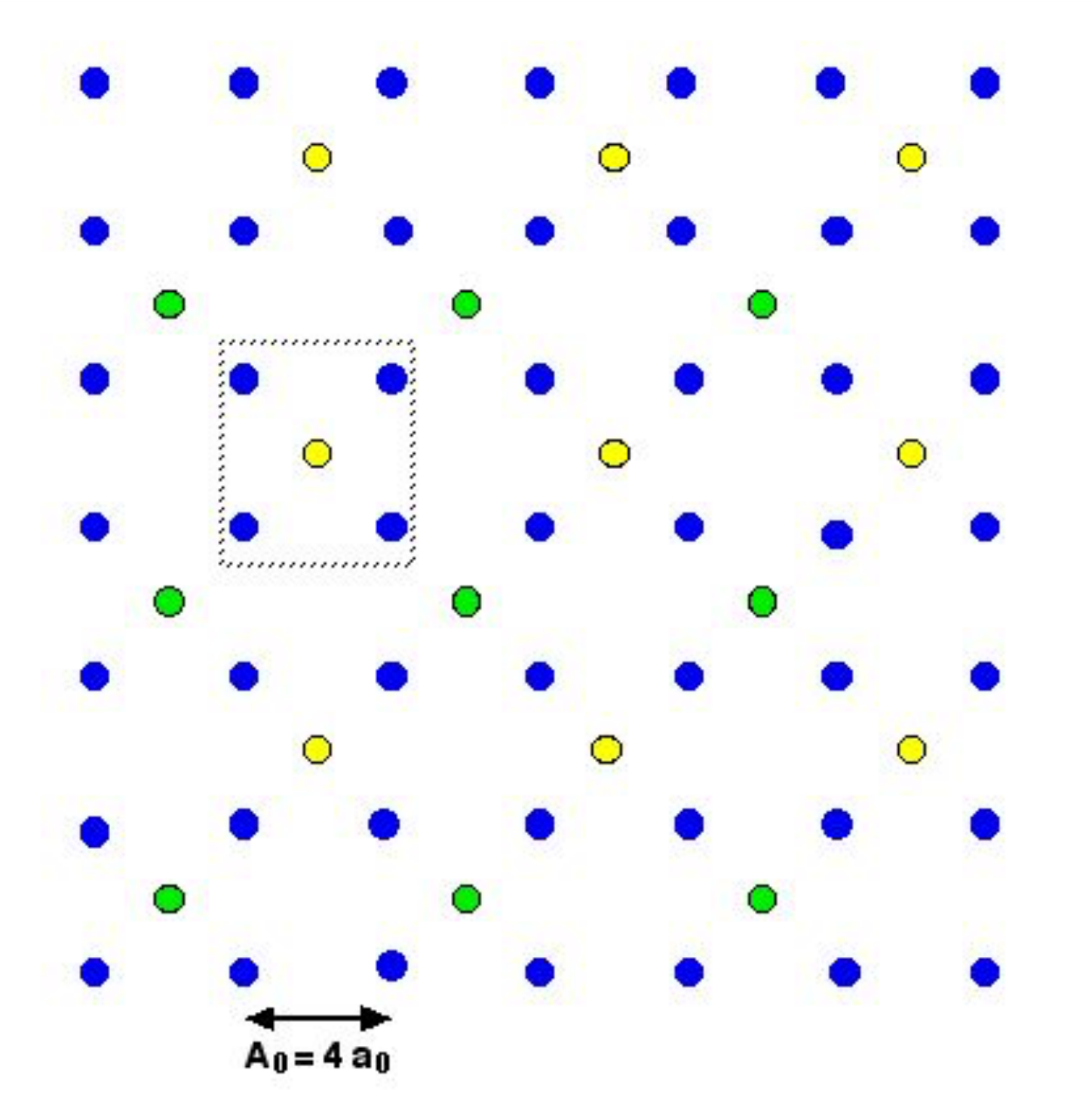}  \footnotesize

\noindent FIG. 4. Commensurable superlattices of $O$ atoms (size exaggerated) in the $CuO_2$ plane (blue), upper $\overline{LaO}$ layer (yellow), and lower $\underline{LaO}$ layer (green). Ions of the host lattice are not shown. The framed rectangle corresponds to Fig. 3.
The same symmetric pattern holds (slightly extended or contracted) for the incommensurable $O$ superlattices when the pseudogap closes in $La_{2-x}Sr_xCuO_4$ and $La_{1.6-x}Nd_{0.4}Sr_xCuO_4$.
\normalsize 
\pagebreak

\end{document}